\documentclass[12pt]{iopart}
\usepackage{graphicx}
\usepackage{iopams}
\usepackage{color}

\def\be{\begin{equation}}
\def\ba{\begin{eqnarray}}
\def\a{\alpha}
\def\b{\beta}

\def\d{\delta}
\def\D{\Delta}

\def\r{\rho}

\def\i{\int}

\def\Tr{\mbox{Tr}}

\def\ee#1{\label{#1}\end{equation}}
\def\ea#1{\label{#1}\end{eqnarray}}
\begin{document}
\title[Fluctuation theorems in driven open quantum
systems]{Fluctuation theorems in driven open quantum systems}  
\author{Peter Talkner, Michele Campisi, Peter H\"anggi}
\address{Institute of Physics, University of Augsburg,
D-86135 Augsburg, Germany}
\ead{peter.talkner@physik.uni-augsburg.de}
\date{\today}
\begin{abstract}

The characteristic function for the joint measurement of the changes
of  two commuting observables upon an external forcing of a quantum
system is derived. In particular, the statistics of the internal
energy, the exchanged heat and the work of a quantum system that {\it weakly}
couples to its environment 
is determined in terms of the energy changes
of the system and the environment due to the action of a classical,
external 
force on the system.  If the system and environment initially
are in a canonical equilibrium, the work performed on the system is
shown to satisfy the Tasaki-Crooks theorem and the Jarzynski equality. 
\end{abstract}
\submitto{Journal of Statistical Mechanics: Theory and Experiment}
\maketitle
\section{Introduction}

About one decade ago Jarzynski proved a quite remarkable equality that
relates free energy differences to average exponentiated work done on
a thermally insulated system that is acted upon by external time
dependent forces  varying according to a specified protocol. This equation,  
now commonly referred to as the Jarzynski Equality, reads \cite{Jarz97}:
\be
\langle e^{-\b w} \rangle = e^{-\b \D F},
\ee{JE}
with $w$ the work, $\D F$ the free energy difference between a
reference equilibrium state of the system at the initial temperature $\b$ with the
force values reached at the end of the force protocol and the truly thermal
initial state. Note that the Jarzynski Equality holds irrespectively
of whether the system ever reaches this reference equilibrium state. 

Crooks \cite{Crooks99} later showed that  Eq. (\ref{JE}) results from
the following work fluctuation theorem: 
\be
\frac{p_{t_f,t_0}(w)}{p_{t_0,t_f}(-w)} = e^{-\b (\D F-w)}
\ee{TC}
that relates the probability density function (pdf) of work $p_{t_f,t_0}(w)$ in the real
forward process proceeding from the time $t_{0}$ until $t_{f}$ to the
pdf of negative work $p_{t_0,t_f}(-w)$ of the mirror 
image process where the time ideally runs backward. 

An important question that naturally arises is whether Eqs. (\ref{JE})
and (\ref{TC}) keep holding in the more realistic situation where the
system remains in thermal contact with its environment while the
forcing protocol is in action.  
In regard to Eq. (\ref{JE}) a positive answer to this question was
already given in 
Ref. \cite{Jarz97} on the basis of classical arguments. 

Moreover, quantum extensions of Eqs. (\ref{JE}) and (\ref{TC}) were
developed, too, first the quantum version of the Jarzynsky equality
for cyclic processes \cite{Kurchan},
and shortly after the Crooks theorem was demonstrated to hold also for
closed quantum systems \cite{tasaki-2000}. A specification of the full
statistics in terms of the characteristic function, i.e. the Fourier
transform of the probability density function (pdf) of work  was
obtained in \cite{TLH}, and a proof of the Crooks relation on the basis of
this characteristic function was given in Ref. \cite{TH}. A
generalization to arbitrary initial states such as the microcanonical
state was obtained in Ref. \cite{THM}. In the latter case a microcanonical Crooks
theorem was proved by means of which the change of the thermodynamic
entropy can be inferred from the statistics of the work \cite{THM,campisi08b}. Illustrative
examples for differently driven quantum harmonic oscillators were
discussed in Refs. \cite{DL,TBH}. An experimental test was proposed in
Ref. \cite{HSKDL}.     

Quantum generalizations of the Jarzynski equality and the Crooks
theorem for open quantum systems have almost exclusively been studied for
systems with Markovian dynamics
\cite{Mukamel03,DeRoeck04,Esposito06,Crooks08}. 
Therefore, so far, only systems with 
weak coupling to their environments have been considered as weak
coupling is implicit in the Markovian assumption
\cite{GWT,T,RHW,HI}
\footnote{In the strong coupling 
  limit the dynamics of a particle's position is described by a classical
  Smoluchowski equation \cite{AGP,MKHTL,CKTM,CKTC} but this equation does
  not provide information about the dynamics of the momentum. }  .

In the present paper we too restrict ourselves to the case of weak
interaction between the system and its environment. We do so because to our best
knowledge presently an unambiguous definition of work and heat of a small open
quantum system is only known in this very case of weak coupling \cite{AN,HB}.
However, neither the dynamics of the system nor of the bath is
restricted otherwise. In
this way we allow for {\em general non-Markovian dynamics} imposed by the bath
and arbitrary force protocols which are neither constrained to be fast
nor slow.

Here we employ the characteristic function approach of
Refs. \cite{TLH,THM} to address the
question regarding the validity of the Eqs. (\ref{JE}) and (\ref{TC})
in quantum systems in weak contact with their environment. 
Hence the focus of this study is on the characteristic function of the
\emph{joint} statistics of simultaneous measurements of system and
environmental energies, 
which are commuting observables. Under the assumption of
weak coupling  the changes of these two energies imposed by the
action of an external force can immediately be related to the changes of
internal energy and the heat exchanged with the environment, or,
equivalently, to work and heat.  
Our central result is the derivation of a fluctuation theorem of the
Tasaki-Crooks type, for the joint pdfs of either internal energy and
heat or, heat and work [see Eq. (\ref{TCEQ})
below]. 
As a corollary of this theorem, we recover the Jarzynsky equation
(\ref{JE}) and the Crooks theorem 
(\ref{TC}) for the marginal pdf of work. This formally proves
that the validity of these relations keeps holding for quantum systems
weakly interacting 
with an environment. 

Previous investigations of the
Jarzynski equality for open Markovian quantum systems \cite{DeRoeck04,Crooks08}
are based on the limit of infinitely many repeated measurements of the 
environment's energy. In contrast, 
the present approach relies on energy measurements of the
system and the environment at the beginning and the end
of the protocol.   
We note that other fluctuation theorems are known in literature
for the heat exchange between two systems which either couple directly
to each other \cite{JW}, or through an intermediate system giving rise to
a heat flow \cite{Saito}. In contrast, here we consider systems which
initially are in thermal equilibrium with their environment and are
then driven out of equilibrium by a classical external force.

The paper is organized as follows. In Sec. \ref{sec:charfun} we define
and obtain the general expression for the joint pdf of
internal energy and heat. In Sec. \ref{sec:fluctheo} the joint
pdf is further evaluated for the initial canonical state
of the \emph{total} system and 
a fluctuation
theorem for work and heat is obtained. Conclusions are drawn in
Sec. \ref{sec:conclusions}. 


\section{\label{sec:charfun}Characteristic functions of work, heat and internal energy} 
We consider a system $S$ that is in weak contact with its environment
$B$. Accordingly, the Hamiltonian $H(t)$ of the total system consists of a
system and an environmental part $H^{S}(t)$ and $H^{B}$, respectively,
and the interaction Hamiltonian $H^{SB}$, i.e.
\be
H(t) = H^{S}(t)+H^{SB} + H^{B} \;,
\ee{H}
where the interaction is small compared to both the system and the
environmental Hamiltonians. We assume that the gauge is fixed in such a
way that the Hamiltonian $H^{S}(t)$ coincides with the system energy despite
its time dependence \cite{Kobe}. This time dependence is caused by an
external change of system parameters according to a prescribed
protocol. Since the system and the
environmental Hamiltonians commute with each other the energies of the
system and of the bath can be measured simultaneously with possible
results $e^{S}_{i}(t)$ and $e^{B}_{\a}$ which are the eigenvalues of
the operators $H^{S}(t)$ and $H^{B}$, respectively. The corresponding 
projection operators  onto the common
eigenfunctions of these operators are denoted by $P_{i,\a}(t)$. 
Hence, the eigenvalues and
eigenprojection operators are determined by the following equations: 
\be
H_{S}(t) P_{i,\a}(t) = e^{S}_{i}(t) P_{i,\a}(t), \quad   H_{B}
P_{i,\a}(t) = e^{B}_{\a} P_{i,\a}(t) \; .
\ee{HSHB}
The projection operators onto the common eigenspaces of two
commuting hermitian operators are hermitian: 
\be
P^{\dagger}_{i,\a}(t) = P_{i,\a}(t)\:,
\ee{Ph}
idempotent and  mutually
orthogonal: 
\be
P_{i,\a}(t)P_{i',\a'}(t) = \d_{i,i'}\d_{\a,\a'} P_{i,\a}(t)\:,
\ee{Pimo}
and complete:  
\be
\sum_{i,\a} P_{i,\a}(t)= 1\!\!1\:,
\ee{P} 
where $1\!\!1$ denotes the unit operator on the total Hilbert space of
system and environment.
The first measurement is performed at the time $t_{0}$
at which the protocol starts to act. At this very time the state of
the total system is
assumed to be given by the density matrix $\r(t_{0})$ which we
will specify later. The joint probability to measure the respective system and
environmental energies $e^{S}_{i}(t_{0})$ and 
$e^{B}_{\a}$, in this state is given by
\be
p_{i,\a} = \Tr P_{i,\a}(t_{0})\r(t_{0})\; ,
\ee{pia}
where $\Tr$ denotes the trace over the total Hilbert space of the
system and the environment. After a measurement with the outcome
$e^{S}_{i}(t_{0}),\;e^{B}_{\a}$ 
the system is found in the initial state projected unto the corresponding
subspace.
This state is given by
\be
\r_{i,\a} = p^{-1}_{i,\a} P_{i,\a}(t_{0}) \r(t_{0}) P_{i,\a}(t_{0}) \; .
\ee{ria}
The second measurement of the system and the environmental energies is
performed at the end of the protocol at time $t_{f}$. By then 
the density matrix of the total system has undergone a 
unitary time evolution in the total Hilbert space to a new state reading
\be
\r_{i,\a}(t_{f}) = U_{t_{f},t_{0}} \r_{i,\a} U^{\dagger}_{t_{f},t_{0}}
\; .
\ee{riat}
The result of the second measurement is characterized by the conditional
pdf of finding energies $e^{S}_{i'}(t_{f})$ and $e^{B}_{\a'}$ 
given that the result of the first measurement were $e^{S}_{i}(t_{f})$
and $e^{B}_{\a}$. This pdf is 
given by 
\be
p_{t_{f},t_{0}}(i',\a'|i,\a) = \Tr P_{i',\a'}(t_{f})
\r_{i,\a}(t_{f})\; .
\ee{piaf}
Consequently, the joint pdf $p_{t_{f},t_{0}}(\D e^{S}, \D
e^{B})$ to measure changes of the system
and the environmental energies $\D e^{S}$ and $\D e^{B}$, respectively, becomes
\ba
p_{t_{f},t_{0}}(\D e^{S}, \D e^{B})& = \sum_{i,i',\a,\a'} \d \left (\D
  e^{S} -(e^{S}_{i'}(t_{f})-e^{S}_{i}(t_{0})  )\right)\nonumber \\
&\qquad \times  \d \left (\D
  e^{B} -(e^{B}_{\a'}-e^{B}_{\a}  )\right)
p_{t_{f},t_{0}}(i',\a'|i,\a) p_{i,\a}\; .
\ea{pdSdB}
At weak coupling between the system and the environment 
the random quantity $\D e^{S}$ determines the change of the internal
energy of the system. 
Then the change of the environmental energy equals the
amount of energy which is exchanged as 
heat $Q = - \D e^{B}$ with the system. 
The small contribution of energy that is possibly
released from or stored in the interaction Hamiltonian is negligibly
small for a system weakly coupled to its environment. Hence, the
joint pdf of internal energy change $E= \D e^{S}$ and
heat exchange  $Q$ becomes
\be \fl
p_{t_{f},t_{0}}(E,Q) = \sum_{i,i',\a.\a'} \d\left (E\!-\!
  (e^{S}_{i'}(t_{f})\!-\!e^{S}_{i}(t_{0}))\right ) \d\left (Q\! +\!
  (e^{B}_{\a'}\!-\!e^{B}_{\a})\right ) p_{t_{f},t_{0}}(i',\a'|i,\a)
p_{i,\a} \; .
\ee{pEQ}
   
The characteristic function $G^{E,Q}_{t_{f},t_{0}}(u,v)$ provides an
equivalent description of the statistics of these energy changes. It
is given by the Fourier transform of 
this pdf, which, due to the presence of the delta functions, can
readily be performed to yield
\ba
G^{E,Q}_{t_{f},t_{0}}(u,v) &= \i dE \i dQ e^{i(u E + v Q)}
p_{t_{f},t_{0}}(E, Q) \nonumber \\ 
&= \sum_{i,i',\a,\a'} e^{i\left( u (e^{S}_{i'}(t_{f})
    -e^{S}_{i}(t_{0})) \right) } e^{-i\left( v (e^{B}_{\a'}
    -e^{B}_{\a}) \right) } p(i',\a'|i,\a) p_{i,\a} \; .
\ea{GS}  
This expression can be further simplified by means of the 
Eqs. (\ref{pia}), (\ref{pdSdB}) and the completeness relation 
(\ref{P}) into the form of a correlation function, reading
\be
G^{E,Q}_{t_{f},t_{0}}(u,v) = \Tr e^{i(u H^{S}_{H}(t_{f}) -v H^{B}_{H}(t_{f}))}
e^{-i(u H^{S}(t_{0}) -v H^{B})} \bar{\r}(t_{0})
\ee{GT}
where the index $H$ denotes the Heisenberg picture of the
corresponding operators , i.e.
\be
H^{S}_{H}(t_{f}) = U^{\dagger}_{t_{f},t_{0}}
H^{S}(t_{f})U_{t_{f},t_{0}}, \quad  H^{B}_{H}(t_{f}) =
U^{\dagger}_{t_{f},t_{0}} H^{B}U_{t_{f},t_{0}} \; .
\ee{HH}
The density matrix $\bar{\r}(t_{0})$ describes the system immediately
after the first measurement. It is given by the projection of the
initial density matrix $\r(t_{0})$ onto the diagonal
states with respect to the eigenbasis of the measured operators. It
hence assumes the form
\be
\bar{\r}(t_{0}) = \sum_{i,\a} P_{i,\a}(t_{0}) \r(t_{0})
P_{i,\a}(t_{0}) \; .
\ee{br}  
In analogy to the case of the measurement of the total energy of an isolated
system the characteristic function for a joint measurement of two
energies is determined by a correlation function of the exponential
operator $(\exp \left [i(u H^{S}(t_{f}) -v H^{B}) \right
])_{H}(t_{f})$ in the Heisenberg picture at the time of the second
measurement $t_{f}$ and a second exponential operator $\exp \left
  [-i(u H^{S}(t_{0}) -v H^{B}) \right]$ taken at the initial time. 
The average is performed with
respect to the initial density matrix projected onto the diagonal
elements with respect to the joint eigenbasis of the two operators of
the first measurement.

In passing we note that in an analogous way as for the internal energy and
the exchanged heat represented by the Hamiltonians of system and
environment, respectively, the characteristic function of the changes $x_{i}$ of any
mutually commuting set of $N$ observables $X^{i}(t)$ can be represented as a
correlation function 
\be
G^{\{x_{i}\}}_{t_{f},t_{0}}(u_{1},u_{2}, \ldots u_{N}) = \Tr e^{i\sum_{i=1}^{N}
  u_{i} X^{i}_{H}(t_{f})} e^{-i\sum_{i=1}^{N}
  u_{i} X^{i}(t_{0})} \bar{\r}_{0}
\ee{Gx}
where $u_{i}$ denotes the Fourier variable conjugate to $x_{i}$,
$X^{i}_{H}(t_{f})$ the observable $X^{i}(t)$  in the Heisenberg
picture and where the density matrix 
$\bar{\r}_{0}$ is given by the projection of the initial density matrix
$\r_{0}$ onto the common diagonal basis of the observables
$X_{i}(t_{0})$, $i=1\ldots N$ as in Eq. (\ref{br}).

Once the statistical properties of internal energy changes and the
exchange of heat are known, the marginal 
distributions of these quantities can equivalently be characterized in terms of
their respective characteristic functions $G^{E}_{t_{f},t_{0}}(u)$ and
$G^{Q}_{t_{f},t_{0}}(v)$; i.e., 
\ba
G^{E}_{t_{f},t_{0}}(u)& = G^{E,Q}_{t_{f},t_{0}}(u,0)\; .\nonumber\\
G^{Q}_{t_{f},t_{0}}(v)& = G^{E,Q}_{t_{f},t_{0}}(0,v) \; .
\ea{GEGQ}  
Moreover, the internal energy change and the heat exchange determine  the
work 
\be
w = E-Q
\ee{w}
performed on the system according to the First Law. Correspondingly, the
joint characteristic function of heat and work,
$G^{Q,w}_{t_{f},t_{0}}(x,y)$, is related to that of internal
energy and heat by
\be
G^{Q,w}_{t_{f},t_{0}}(x,y) = G^{E,Q}_{t_{f},t_{0}}(y,x-y) \; .
\ee{GQw}
Then, the marginal characteristic function of work performed on
the system becomes
\be
G^{w}_{t_{f},t_{0}}(z) = G^{E,Q}_{t_{f},t_{0}}(z,-z) \; .
\ee{Gw}    
 
So far we have not yet specified the initial density matrix
$\r(t_{0})$.
In the next section we consider the particularly relevant case of a
canonical state of the total system at a given temperature.

\section{\label{sec:fluctheo}Fluctuation theorem for work and heat}
We now assume that the total system consisting of the considered system
and its environment is initially, i.e. at $t=t_{0}$, in a
thermodynamical equilibrium at inverse temperature $\b$ and is
consequently described by the Gibbs state
\be
\r_{\b}(t_{0}) = Z^{-1}(t_{0}) e^{-\b(H^{S}(t_{0})+H^{B} +H^{SB})} \; ,
\ee{rbeta}
where  
\be
Z(t_{0}) = \Tr e^{-\b(H^{S}(t_{0})+H^{B} +H^{SB})}  
\ee{Z}
denotes the partition function. In the particular case of weak
coupling between the system and its environment a perturbation
expansion of the density matrix with respect to the interaction
Hamiltonian yields up to first order  
\be \fl
\r_{\b}(t_{0})\approx \r^{0}(t_{0}) - Z_{S}^{-1}(t_{0}) Z_{B}^{-1}
\i_{0}^{\b} d \b'
  e^{-(\b-\b')(H^{S}(t_{0})+H^{B})} \d H^{SB}
  e^{-\b'(H^{S}(t_{0})+H^{B})} \; ,
\ee{rb1}  
where
\ba \label{ZS}
Z^{S}(t_0) &= \Tr_{S}\: e^{-\b H^{S}(t_{0})}\\ 
Z^{B} &= \Tr_{B}\: e^{-\b H^{S}}\; .
\ea{ZB}
Here $\Tr_{S}$ and $\Tr_{B}$ denote the traces over the system and
the environmental Hilbert spaces, respectively. The operator $\d H^{SB} = H^{SB} -
\langle H^{SB} \rangle_{0}$  specifies the deviation of the
interaction from its expectation value with respect to the factorizing
state 
\be
\r^{0}(t_{0}) =Z_{S}^{-1}(t_{0}) Z_{B}^{-1} e^{-\b(H^{S}(t_{0}) +
  H^{B})}\; .
\ee{r0}
In order to determine the density matrix $\bar{\r}(t_{0})$,
Eq. (\ref{br}), $\r_{\b}(t_{0})$ has to be projected
onto the diagonal elements with respect 
to the joint energy
eigenbasis of $H^{S}(t_{0})$ and $H^{B}$. This 
projection leaves the unperturbed part
$\r^{0}(t_{0})$ of the initial density matrix (\ref{rb1}) unchanged. The first
order correction on the right hand side of Eq. (\ref{rb1}) vanishes in
all cases when the 
interaction between system and environment contains only off-diagonal
terms with respect to the unperturbed energy basis as is the case
e.g. for 
the spin-Bose model \cite{Leggett87}, or the Caldeira-Leggett model
\cite{CaldeiraLeggett81} \footnote{If the interaction
  contains diagonal terms, these can safely be added to the
  system Hamiltonian, which then would be  re-defined
  accordingly.}. In all these cases 
the corrections to the factorizing density matrix $\r^{0}(t_{0})$ are
at least of second order in to the system-environment interaction. Therefore
they can safely be neglected in the limit of weak coupling such that the
diagonal projection of $\r_{\b}(t_{0})$ leads to the factorizing state
$\r^{0}(t)$,
i.e.
\be
\bar{\r}_{\b}(t_{0}) = \r^{0}(t_{0}) \equiv Z^{-1}_{S}(t_{0})
Z^{-1}_{B} e^{-\b (H^{S}(t_{0}) +H^{B})} \; .
\ee{brr0}
We emphasize that this holds under the conditions of weak interaction
between the system and its environment. 

With this initial state the characteristic function for the statistics
of internal energy and heat becomes
\be \fl
G^{E,Q}_{t_{f},t_{0}}(u,v) = Z^{-1}_{S}(t_{0})Z^{-1}_{B} \Tr \:
e^{i\left ( uH^{S}_{H}(t_{f}) -vH^{B}_{H}(t_{f}) \right)} 
e^{-i\left ( uH^{S}(t_{0}) -vH^{B} \right)} e^{-\b \left
    (H^{S}(t_{0})+H^{B} \right )} \; .
\ee{GEQS}
Analogous to the characteristic function of work performed on a
closed system initially in a canonical state \cite{TH}, 
this expression can be continued to an analytic function
of the variables $u$ and $v$ on stripes $\mathcal{S}_{u}=\left
  \{u=u'+iu''|u'\in \mathbb{R}, 0\leq u'' \leq \b \right \}$  and 
$\mathcal{S}_{v}=\left
  \{v=v'+iv''|v'\in \mathbb{R}, -\b\leq v'' \leq 0 \right \}$ of
the complex plane. The proof is based on the fact that 
$\exp{iu H^{S}(t)}$ and $\exp{-iv H_{B}}$ are trace class operators
which are analytical functions of
$u$ and $v$ in the specified strips \cite{KMS}. 
Setting $u=\bar{u}+i\b$ and $v=\bar{v}+i\b$ one finds 
\ba
Z^{S}(t_{0}) G^{E,Q}_{t_{f},t_{0}}(u,v) &= Z^{S}(t_{f})
G^{E,Q}_{t_{0},t_{f}}(-\bar{u},-\bar{v}) \nonumber \\
&=  Z^{S}(t_{f})
G^{E,Q}_{t_{0},t_{f}}(-u+i\b,-v-i\b) \; ,
\ea{GTC}
where $G^{E,Q}_{t_{0},t_{f}}(u,v)$ is the characteristic function of
the 
internal energy change and heat transfer for a fictitious process that
runs under the action of the time reversed process backward in time. 
The proof is analogous to the case of a closed system initially in a
canonical state \cite{TH}. Applying the inverse Fourier transform with
respect to both variable $u$ and $v$ one obtains a Tasaki-Crooks type
expression \cite{Crooks99,tasaki-2000}. It establishes a connection
between  the joint pdf of internal energy 
change and exchanged heat for the original process to the
corresponding quantity of the time reversed process. This fluctuation
theorem reads accordingly
\be 
\frac{p_{t_{f},t_{0}}(E,Q)}{p_{t_{0},t_{f}}(-E,-Q)} =
\frac{Z^{S}(t_{f})}{Z^{S}(t_{0})}\: e^{\b(E-Q)} = e^{-\b(\D F -E +Q)}
\; ,
\ee{TCEQ}
where $\D F = - \b^{-1} \ln [Z^{S}(t_{f})/Z^{S}(t_{0})]$ 
denotes the free energy difference between the system
in the canonical state with the parameter values at
$t_{f}$  and in the initial state at $t_{0}$. 

\subsection{Nonequilibrium free-energy/work relations}
Replacing the internal
energy by the work performed on the system we obtain 
\be
Z^{S}(t_{0}) G^{Q,w}_{t_{f},t_{0}}(x,y) = Z^{S}(t_{f})
G^{Q,w}_{t_{0},t_{f}}(-x-y,-y+i\b) \; , 
\ee{FTGQw}
or equivalently
\be
\frac{p^{Q,w}_{t_{f},t_{0}}(Q,w)}{p^{Q,w}_{t_{0},t_{f}}(-Q,-w)} =
e^{-\b(\D F -w)} \; .
\ee{QwTC}
Here $p^{Q,w}_{t_{f},t_{0}}(Q,w)= p_{t_{f},t_{0}}(Q+w,Q)$  
denotes the joint pdf of heat and work 
corresponding to the characteristic function
$G^{Q,w}_{t_{f},t_{0}}(x,y)$ defined in Eq. (\ref{GQw}). 
Integrating over all possible values of the exchanged heat one obtains
the Tasaki-Crooks theorem for an open system which couples weakly to
its environment. It reads
\be
\frac{p^w_{t_{f},t_{0}}(w)}{p^w_{t_{0},t_{f}}(-w)} = e^{-\b(\D F -w)}
\; ,
\ee{wTC}
where $p^w_{t_{f},t_{0}}(w) = \i dQ\: p^{Q,w}_{t_{f},t_{0}}(Q,w) $ denotes the
marginal pdf of work.
From the Tasaki-Crooks theorem Jarzynski's work theorem follows
immediately saying that the mean value of the exponentiated work
performed on an open system in weak contact with its environment
coincides with the exponentiated free energy difference \cite{Jarz97}, i.e.,
\be
\langle e^{-\b w} \rangle = e^{-\b \D F} \; .
\ee{J}

\subsection{\label{CondProbs}Nonequilibrium relations for conditional and marginal probabilities}
From Eq. (\ref{TCEQ}) one can obtain nonequilibrium relations for the
marginal distribution of heat  
\be
p^Q_{t_{f},t_{0}}(Q) = \i dw p^{Q,w}_{t_{f},t_{0}}(Q,w),
\ee{pQ}
for the pdf of work $w$ under the condition that the heat
$Q$ is measured, i.e.,:
\be
p_{t_{f},t_{0}}(w|Q) =
\frac{p^{Q,w}_{t_{f},t_{0}}(Q,w)}{p^Q_{t_{f},t_{0}}(Q)} \; ,
\ee{pwQ}
and for the pdf of heat $Q$ under the condition that the work $w$ is
measured, i.e.:
\be
p_{t_{f},t_{0}}(Q|w) =
\frac{p^{Q,w}_{t_{f},t_{0}}(Q,w)}{p^w_{t_{f},t_{0}}(w)} \; ,
\ee{pQcw}
A straightforward calculation yields:
\be
\frac{p^Q_{t_{0},t_{f}}(-Q)}{p^Q_{t_{f},t_{0}}(Q)} = e^{\beta \D F}
\langle e^{-\b w}|Q\rangle \; ,
\ee{pQQ}
where the symbol $\langle \cdot |Q\rangle$ denotes average over $p_{t_{f},t_{0}}(w|Q)$.

On the other hand one finds the following relation
\be
\frac{p_{t_{f},t_{0}}(Q|w)}{p_{t_{0},t_{f}}(-Q|-w)} =1.
\ee{pQw}
for the conditional pdf of heat $Q$ given the work $w$, in the
forward and backward processes.

\subsection{Nonequilibrium equalities for exponentiated internal energy and heat}

By noting that 
\be
\langle e^{\b Q}\rangle_Q = \i dQ p_{t_f,t_0}^{Q}(Q) e^{\b Q} = G_{t_f,t_0}^{Q}(-i \b)
\ee{eQ}
and 
\be
\langle e^{-\b E}\rangle_E = \i dE p_{t_f,t_0}^{E}(E) e^{-\b E} = G_{t_f,t_0}^{E}(i \b),
\ee{eE} 
and using Eqs. (\ref{GEGQ}) and (\ref{GEQS}) one finds two
nonequilibrium equalities for heat and internal energy: 
\be
\langle e^{\b Q}\rangle_Q = \frac{\Tr e^{-\b H^S(t_0)} e^{-\b H_H^B(t_f)}} {Z_S(t_0)Z_B}
\ee{JEQ}
\be
\langle e^{-\b E}\rangle_E = \frac{\Tr e^{-\b H_H^S(t_f)} e^{-\b H^B}} {Z_S(t_0)Z_B}.
\ee{JEE}
In clear contrast to the Jarzynski equality, the average of the
exponentiated heat as well as that of the exponentiated energy do depend
on the details of the protocol, i.e. these relations do not 
depend only on equilibrium properties
of the system.

\section{\label{sec:conclusions}Conclusions}
We proved that the Tasaki-Crooks fluctuation theorem and the
Jarzyinski equality hold in open quantum systems that are weakly coupled to
their environments. 
This was accomplished by introducing the characteristic function for
the probability of joint measurements of system and environmental
energies, which are commuting observables. This characteristic functions
was further evaluated for  canonical initial states. Our
approach rests on two basic assumptions: First, the 
total system made of the system of interest and its environment
is governed by Hamiltonian dynamics and second, the coupling between
system and its environment is assumed to
be weak. The first assumption leads to a unitary time evolution of the
total system. The second assumption has two important consequences: It
allows one to determine the internal energy from the
Hamiltonian of the system and the heat from the Hamiltonian of the
environment. The work performed on the system then follows by means of the First
Law. The contribution of the interaction Hamiltonian is neglected in
the definitions of these three energies. The second consequence of the
weak coupling assumption is the factorization of the 
state of the total system immediately after the first measurements of
energies into a product of the Gibbs states of the
system and the environment as if they were uncoupled up to second
order corrections in the system environment interaction. In order for
these corrections to be small for weak but nonzero coupling, the
temperature of the initial state must not be too low. In
contrast, we formally allowed for the exact time evolution such that
fast as well as slow protocols can be described adequately.

By an inverse Fourier transformation the joint statistics of internal energy
and heat (defined as the energy ceded to the environment) was shown to
obey a Tasaki-Crooks type fluctuation theorem. From this we
derived the Tasaki-Crooks fluctuation theorem and the Jarzynski
equality, for a quantum system that weakly couples to its
environment \cite{TH}. 
Further nonequilibrium relations for the marginal and conditional
pdfs of work given heat and heat given work were derived.

In the present paper we restricted ourselves to the case where the
external forces directly influence the system Hamiltonian but
leave the interaction and the environment Hamiltonians unchanged. Our
approach can be generalized to more complex situations where the forces
immediately influence the system plus environment. In this general
case the Tasaki-Crooks theorem, Eq. (\ref{QwTC}), continues to hold in a
slightly modified form: Also the
change of the free energy of the environment has to be taken into
account, i.e. $\D F$ has to be replaced by $\D F +\D F_{B}$, where $\D
F_{B}$ denotes the change in free energy of the environment due to the
parameter change of the environmental Hamiltonian. In those special cases
where only the system and the interaction Hamiltonians are influenced by the
external forcing, the form of the Tasaki-Crooks theorem as given by
Eq. (\ref{QwTC}) stays unchanged. As a consequence the Jarzynski
equality remains valid for forces changing the system and its
interaction 
to the environment but leaves the Hamiltonian of the
environment unchanged.   

For a system that strongly couples to its environment it frequently is 
possible to ``dress'' the system with excitations of its
environment such that the dressed system only weakly interacts with its
properly redefined environment. Quasi-particles in solid state
physics are typical examples of this type of systems. An external
force acting  on the original system will also be felt by the dressed
system. For the work exerted on the dressed system and the heat
exchanged with the properly redefined environment 
all conditions for the presented theory apply and therefore  
the Tasaki-Crooks theorem and
consequently the Jarzynski equality remain valid for such dressed
system. Note, however, that for the original ``bare'' system, as for
other strongly coupled 
systems, these relations cannot be expected to hold.
\subsection*{Acknowledgments}
Financial support by the German Excellence Initiative via the {\it
  Nanosystems Initiative Munich} (NIM) and the Volkswagen
Foundation (project I/80424) is greatfully acknowledged. 
\section*{References} 
\bibliographystyle{unsrt} 

\begin{thebibliography}{10}

\bibitem{Jarz97}
Jarzynski C,
\newblock  {\em Nonequilibrium equality for free energy differences}, 1997
\newblock {\em Phys. Rev. Lett.} {\bf 78} 2690
\bibitem{Crooks99}
Crooks G E,
\newblock {\em Entropy production fluctuation theorem and the nonequilibrium work
  relation for free energy differences}, 1999
\newblock {\em Phys. Rev. E} {\bf 60} 2721

\bibitem{Kurchan}
Kurchan G,
\newblock {\em A quantum fluctuation theorem}, 2000 {\em e-print cond-mat/0007360}


\bibitem{tasaki-2000}
Tasaki H,
\newblock {\em Jarzynski relations for quantum systems and some applications}, 2000
\newblock {\em preprint cond-mat/0009244}

\bibitem{TLH}
Talkner P, Lutz E and H\"{a}nggi P,
\newblock {\em Fluctuation theorems: Work is not an observable}, 2007
\newblock {\em Phys. Rev. E} {\bf 75} 050102

\bibitem{TH}
Talkner P and H\"{a}nggi P,
\newblock {\em The {T}asaki-{C}rooks quantum fluctuation theorem}, 2007
\newblock {\em J. Phys. A: Math. Theor.}
  {\bf 40} F569--F571

\bibitem{THM}
Talkner P, H\"{a}nggi P and Morillo M,
\newblock {\em Microcanonical quantum fluctuation theorems}, 2008
\newblock {\em Phys. Rev. E} {\bf 77} 05113

\bibitem{campisi08b}
Campisi M,
\newblock {\em Complementary expressions for the entropy-from-work theorem}, 2008
\newblock {\em Phys. Rev. E} {\bf 78} 012102

\bibitem{DL}
Deffner S and Lutz E,
\newblock {\em Nonequilibrium work distribution of a quantum harmonic oscillator}, 2008
\newblock {\em Phys. Rev. E} {\bf 77} 021128

\bibitem{TBH}
Talkner P, Burada P S and H\"anggi P,
\newblock {\em Statistics of work performed on a forced quantum oscillator}, 2008
\newblock {\em Phys. Rev. E} {\bf 78} 011115

\bibitem{HSKDL}
Huber G, Schmidt-Kaler F, Deffner S and Lutz E,
\newblock {\em Employing trapped cold ions to verify the quantum {J}arzynski equality}, 2008
\newblock {\em Phys. Rev. Lett.} {\bf 101} 070403

\bibitem{Mukamel03}
Mukamel S,
\newblock {\em Quantum extension of the Jarzynski relation: Analogy with stochastic
  dephasing}, 2003
\newblock {\em Phys. Rev. Lett.} {\bf 90} 170604

\bibitem{DeRoeck04}
De~Roeck W and Maes C,
\newblock {\em Quantum version of free-energy/irreversible-work relations}, 2004
\newblock {\em Phys. Rev. E} {\bf 69} 026115

\bibitem{Esposito06}
Esposito M and Mukamel S,
\newblock {\em Fluctuation theorems for quantum master equations}, 2006
\newblock {\em Phys. Rev. E} {\bf 73} 046129

\bibitem{Crooks08} Crooks G E, {\em On the {J}arzynski relation for
    dissipative quantum dynamics}, 2008 {\em J. Stat. Mech.: Theor. Exp.} P10023

\bibitem{GWT}
Grabert H, Weiss U and Talkner P,
\newblock {\em Quantum theory of the damped harmonic oscillator}, 1984
\newblock {\em Z. Phys. B} {\bf 55} 87--94

\bibitem{T}
Talkner P,
\newblock {\em The failure of the quantum regression hypothesis}, 1986
\newblock {\em Ann. Phys. (NY)} {\bf 167} 390

\bibitem{RHW} Riseborough P, H\"anggi P and Weiss U, {\em Exact
    results for a damped quantum mechanical oscillator}, 1985 {\em Phys. Rev. A}
  {\bf 31} 471--478

\bibitem{HI}H\"anggi H and Ingold G I, {\em Fundamental aspects
    of quantum {B}rownian motion}, 2005 {\em Chaos} {\bf 15} 026105


\bibitem{AGP}
Ankerhold J, Pechukas P and Grabert H,
\newblock {\em Strong friction limit in quantum mechanics: The quantum {S}moluchowski
  equation}, 2001
\newblock {\em Phys. Rev. Lett.} {\bf 87} 086802

\bibitem{MKHTL}
Machura L, Kostur M, H\"anggi P, Talkner P and \L{}uczka J,
\newblock {\em Consistent description of quantum Brownian motors operating at strong
  friction}, 2004
\newblock {\em Phys. Rev. E} {\bf 70} 031107

\bibitem{CKTM}
Coffey W T, Kalmykov Y P, Titov S V, and Mulligan B~P,
\newblock {\em Semiclassical Kramers and Smoluchowski equations for the
  Brownian motion of a particle in an external potential}, 2007
\newblock {\em J. Phys. A: Math. Theor.} {\bf 40} F91--F98

\bibitem{CKTC} Coffey W T, Kalmykov Y P. Titov S V and Cleary L, 
  {\em Smoluchowski equation approach for quantum {B}rownian motion in
  a tilted periodic potential}, 2008 {\em Phys. Rev. E} {\bf 78} 031114

\bibitem{AN} Allahverdyan A E and Nieuwenhuizen T M, {\em
    Fluctuations of work from quantum subensembles: {T}he case against
  quantum work-fluctuation theorems}, 2005 {\em Phys. Rev. E} {\bf 71} 066102

\bibitem{HB} H\"orhammer C and B\"uttner H, {\em Information and
    entropy in quantum {B}rownian motion: {T}hermodynamic entropy
    versus von Neumann entropy} 2007 , {\em e-print arXiv:0710.1716}  

\bibitem{JW}
Jarzynski C and W\'ojcik D,
\newblock {\em Classical and quantum fluctuation theorems for heat exchange}, 2004
\newblock {\em Phys. Rev. Lett.} {\bf 92} 230602

\bibitem{Saito}
Saito K and Dhar A,
\newblock {\em Fluctuation theorem in quantum heat conduction}, 2007
\newblock {\em Phys. Rev. Lett.} {\bf 99} 180601

\bibitem{Kobe}
Kobe D H,
\newblock {\em Gauge-invariant classical Hamiltonian formulation of the
  electrodynamics of nonrelativistic particles}, 1981
\newblock {\em Am. J. Phys.} {\bf 49} 581--588

\bibitem{Leggett87}
Leggett A~J, Chakravarty S, Dorsey A T, Fisher M P A, Garg A,
  and Zwerger W.
\newblock {\em Dynamics of the dissipative two-state system}, 1987
\newblock {\em Rev. Mod. Phys.} {\bf 59} 1--85

\bibitem{CaldeiraLeggett81}
Caldeira A O and Leggett A J,
\newblock {\em Influence of dissipation on quantum tunneling in macroscopic systems}, 1981
\newblock {\em Phys. Rev. Lett.} {\bf 46} 211--214

\bibitem{KMS}
Haag R, Hugenholtz N M and Winnink M,
\newblock {\em On the equilibrium states in quantum statistical mechanics}, 1967
\newblock {\em Commun. Math. Phys.} {\bf 5} 215--236

\end{thebibliography}

\end{document}